\documentclass{article}
\usepackage[utf8]{inputenc}
\usepackage{amsmath}
\usepackage{amsfonts}
\usepackage{verbatim}
\usepackage{bm}
\usepackage{xcolor}
\usepackage{graphicx}
\usepackage{subcaption}
\usepackage{mwe}
\usepackage{authblk}

\newtheorem{theorem}{Theorem}

\newtheorem{corollary}[theorem]{Corollary}

\newtheorem{lemma}[theorem]{Lemma}

\newenvironment{proof}[1][Proof]{\textbf{#1.} }{\ \rule{0.5em}{0.5em}}

\def\LM{\mbox{LM}}

\usepackage{natbib}
\setcitestyle{authoryear,open={(},close={)}} 

\usepackage[title]{appendix}

\title{A multidimensional objective prior distribution from a scoring rule.}
\author[1]{Isadora Antoniano-Villalobos}
\author[2]{Cristiano Villa}
\author[3]{Stephen G. Walker}
\affil[1]{Ca' Foscari University of Venice, Italy}
\affil[2]{Duke-Kunshan University, Suzhou, China}
\affil[3]{University of Texas at Austin, USA}
\date{}
\begin{document}

\maketitle
\begin{abstract}
\noindent The construction of objective priors is, at best, challenging for multidimensional parameter spaces. A common practice is to assume independence and set up the joint prior as the product of marginal distributions obtained via ``standard'' objective methods, such as Jeffreys or reference priors. However, the assumption of independence a priori is not always reasonable, and whether it can be viewed as strictly objective is still open to discussion. In this paper, by extending a previously proposed objective approach based on scoring rules for the one dimensional case, we propose a novel objective prior for multidimensional parameter spaces which yields a dependence structure. The proposed prior has the appealing property of being proper and does not depend on the chosen model; only on the parameter space considered.
\end{abstract}

\section{Introduction}\label{sc_introduction}
The derivation of objective priors  is known to be  challenging when it comes to multidimensional parameter spaces. A possible solution is to assume prior independence and set up the joint prior as the product of the marginals. 
If one wishes to assign an objective prior to a parameter vector $\pmb\theta=(\theta_1,\ldots,\theta_d)$, the independent prior idea would be to have
$$\pi(\pmb\theta) = \pi_1(\theta_1)\times\cdots\times\pi_d(\theta_d),$$
where $\pi_j(\theta_j)$, for $j=1,\ldots,p$, would be obtained using any ``standard'' objective methods, such as Jeffreys prior \citep{Jeffreys61} or the reference prior \citep{Bernardo1979}. However, as models become ever larger, the idea of placing independent priors on all parameters, which are derived from the model, or using Jeffreys or the reference priors, is too problematic; see \cite{Consonni2018}. While the above implementation of objective priors for multidimensional parameter spaces may be advisable in some circumstances, it is important to take into consideration that the assumption of independence {\em a priori} is not always a satisfactory default choice. It can be hardly credible to assign a strategy to be objective due to the lack of a better choice such as a specific justification.

In this paper we concentrate on proposing a prior for multidimensional parameter spaces based on objective considerations which extend the work of \cite{WV2021}.

The review of the available objective approaches to deal with multidimensional parameter spaces is quite straightforward. Jeffreys himself \citep{Jeffreys61} acknowledged the limitation of obtaining a prior distribution for multidimensional parameter spaces employing his approach (i.e. \emph{Jeffreys rule prior}) and proposed to use the a priori independence  (i.e. \emph{independent Jeffreys prior}). See also the discussion in \cite{KW1999}. 
As for the class of reference priors, while a formal definition for the unidimensional case is available, \citep{BBS2009}, the extension to any parameter space has not yet been provided. There are {\em ad hoc} solutions for some particular cases, such as how to deal with nuisance parameters, see \citep{Liseo93}, but these do not provide any clear indication on how to proceed when the parameters cannot be considered nuisance. Furthermore, quite often the resulting prior distribution may be different depending on the ``order of importance'' assigned to the parameters of interest.

Besides the aforementioned obvious limitation, one aspect that has to be taken into consideration when dealing with objective priors, is that they are typically improper. While it is well known that this does not constitute an issue {\em per se} as long as the corresponding posterior is proper, there is still the problem that posterior properness would need to be verified. For multidimensional parameter spaces, the tediousness and difficulty of this task is evident. We refer the reader to \cite{KW1999} for a discussion.

The idea of this paper is to propose a novel objective prior for multidimensional parameter spaces that has the appealing property of being proper. Naturally, this prior will not require the verification of posterior properness and, more importantly, it can be employed in all the scenarios where an improper prior distribution is not suitable, for example when dealing with Bayes Factors. The idea builds on the procedure presented in \cite{WV2021}, where the connection between information, divergence and scoring rules is exploited to obtain an objective prior distribution. In fact, it assumes that the score function is constant, meaning that we do not favour any particular parameter value with respect to others. It also has the property of minimizing a measure of information. A key aspect of the prior is that it does not depend on the chosen model, but only on the parameter space. 

The paper is organised as follows. In Section \ref{sc_priorderivation}, we present the derivation of the prior distribution for the case $d=2$ and show how to derive the general multidimensional case. In Section 3 we illustrate the implementation of the proposed prior distribution for both simulated and real data; the examples we discuss include single sample results as well as frequentist analysis on repeated samples. Finally, Section \ref{sc_discussion} concludes with a brief discussion and some remarks.

\section{Prior derivation}\label{sc_priorderivation}
We start by reviewing the idea from \cite{WV2021}. This is useful, as the extension to the multidimensional case requires more complex calculations but the principles remain largely the same. 
Consider the general relationship between divergence, information and score; that is, for densities $p$ and $q$
\begin{equation}\label{eq:divergencerelationship}
    D(p,q) = I(p) + \int p\,S(q),
\end{equation}
where $D$ denotes a divergence between the two density functions, $I(p)=-\int p\,S(p)$ is the corresponding information, and $S(q)$ is the score function. For a one dimensional parameter, \cite{WV2021} proceeded by taking a Bregman divergence \citep{BREGMAN1967},
$D(p,q)=\int B_\phi(p,q)$, where 
$$B_\phi(p,q)=\phi(p)-\phi(q)-\frac{\partial \phi}{\partial q}(p-q)-\frac{\partial \phi}{\partial q'}(p'-q'),$$
for some convex function $\phi(q,q')$. For a local score function and to achieve the form (\ref{eq:divergencerelationship}) it is necessary to set
$$\phi(q,q')=q\frac{\partial \phi}{\partial q}+q'\frac{\partial \phi}{\partial q'},$$
which is guaranteed through the choice
$\phi(u,v)=u\,\alpha(v/u)$, for some convex function $\alpha$. Then, setting the score function to be 0, results in the objective prior $q$ being the solution to
$$\frac{\partial \phi}{\partial q}=\frac{d}{dx}\,\frac{\partial \phi}{\partial q'}.$$
When the convex function $\alpha$ is taken to be ``simple'' and of the form $\alpha(u)=u^{-(k+1)}$.
The result is
$$q(x)=\frac{k\,a^k}{(a+x)^{k+1}},$$
for some $k\geq 1$ and $a>0$ (details in Appendix \ref{sec:details}). This is a member of the Lomax family of density functions, which we denote $\LM(a,k)$. We will not reduce this further at this point by specifying $a$ and $k$, save to say that $a=1$ and $k=1$ are well motivated choices with respect to invariance and minimizing information, respectively.
To elaborate on these points, if we now undertake the transform $y=1/x$, then 
$$q(y)=\frac{k\,a^k\,y^{k-1}}{(1+ay)^{k+1}},$$
which only matches $q(x)$ when $a=k=1$. Further, after setting $a=1$, which is done without loss of generality for what follows, consider the negative entropy 
$$I(k)=\log k+(k+1)\int \log(1+x)\,\frac{k}{(1+x)^{k+1}}\,dx.$$
The integral is $1/k$, so $I(k)=\log k+1+1/k$, which is minimized at $k=1$.

\subsection{Dimension $2$}\label{sc_bidimensional}
To simplify the notation, we use $x$ and $y$ to represent the two arguments of the prior $q$.
Before going through the scoring rule idea, we introduce the solution and some appealing properties of the multivariate Lomax distribution. 

If independence between $X$ and $Y$ is not assumed a priori, the conditional prior distribution should also take the form of a Lomax distribution, as it is one dimensional and should adhere to the ideas  discussed in the previous section; hence, $\pi(y\mid x)$ will be a Lomax with parameters $a(x)$ and $k(x)$. If we have that, marginally, the prior $\pi(x)$ for $X$ is $\LM(a,k)$, and we take $a(x)=a+x$ and $k(x)=k+1$, then
\begin{equation}\label{BivLom}
    \pi(x,y)=\frac{(k+1)\,(a+x)^{k+1}}{(a+x+y)^{k+2}}\frac{k\,a^k}{\left(a+ x\right)^{k+1}}=
    \frac{k(k+1)a^k}{(a+x+y)^{k+2}},
\end{equation}
which is a bivariate Lomax density function. Moreover, both the marginals are $\LM(a,k)$.

We now consider the bivariate version of the Bregman divergence 
$$D(\bm{p},\bm{q}) = \int_\Omega B_\phi(\bm{p},\bm{q})\,dx\,dy,$$
where the integral is intended over the space $(x,y)\in\Omega\subseteq\mathbb{R}^2$ and, without loss of generality, we assume for now $\Omega=(0,\infty)^2$. Define
$$\bm{p}(x,y)=(p(x,y),p_x(x,y),p_y(x,y)),$$ 
and similarly for $\bm{q}$, where for the bivariate function $f(x,y)$, we denote
$f_x=\partial f/\partial x$, and similarly for $f_y$. 
In detail, we have
\begin{equation}\label{ep:bregman}
    B_{\phi}(\bm{p},\bm{q}) = \phi(\bm{p})-\phi(\bm{q})-\frac{\partial\phi}{\partial q}(p-q)-\frac{\partial\phi}{\partial q_x}(p_x-q_x)-\frac{\partial\phi}{\partial q_y}(p_y-q_y).
\end{equation}
Rearranging, and using integration by parts with assumptions on the densities vanishing at end points, the integral of (\ref{ep:bregman}) becomes
$$\int\left\{\phi(\bm{p})+p\left[-\frac{\partial\phi}{\partial q}+\frac{d}{dx}\frac{\partial\phi}{\partial q_x}+\frac{d}{dy}\frac{\partial\phi}{\partial q_y}\right] - \left[\phi(\bm{q})-q\frac{\partial\phi}{\partial q}-q_x\frac{\partial\phi}{\partial q_x}-q_y\frac{\partial\phi}{\partial q_y}\right]\right\}.$$
To obtain the decomposition of equation \eqref{eq:divergencerelationship}, we need
\begin{equation}\label{eq:phisolution}
\phi(\bm{q}) = q\frac{\partial\phi}{\partial q} + q_x\frac{\partial\phi}{\partial qx} + q_y\frac{\partial\phi}{\partial q_y},
\end{equation}
yielding the score function
$$S(\bm{q}) = -\frac{\partial\phi}{\partial q} + \frac{d}{dx}\frac{\partial\phi}{\partial q_x} + \frac{d}{dy}\frac{\partial\phi}{\partial q_y}.$$
Following the reasoning outlined in \cite{WV2021}, we propose
\begin{equation}\label{eq:proposedphi}
\phi(q,q_x,q_y) = q\,\alpha\left(\frac{q_x}{q},\frac{q_y}{q}\right),
\end{equation}
for some convex function $\alpha(u,v)$. In the absence of information, it is reasonable to assume that $x$ and $y$ carry the same importance for both the definition of the Bergman divergence and of the scoring rule. This is guaranteed by the symmetry condition $\alpha(u,v)=\alpha(v,u)$.
We can also motivate the additive version of the multidimensional case based on the notion of adding dimensions, hence we take
$$\alpha(u,v)=\frac{1}{u^{k+2}}+\frac{1}{v^{k+2}}.$$
Now, setting $S(\bm{q})=0$ (which we recall being our ``objective'' argument) results in the solution $q(x,y)=\pi(x,y)$ given in (\ref{BivLom}).

\subsection{General multidimensional case}

The extension to the $d$-dimensional case, with $d>2$, is relatively straightforward, using the following convex function:
$$\alpha(u_1,\ldots,u_d)=\sum_{j=1}^d \frac{1}{u_j^{k+d}}.$$If we apply the same strategy as outlined above, the solution results in
$$q(x_1,\ldots,x_d)=\frac{k(k+1)\ldots (k+d-1)\,a^k}{(a+x_1+\cdots +x_d)^{k+d}},$$
which is the density of a multivariate Lomax, denoted $\LM(d,k,a)$. All the marginal and the conditional density functions also belong to the Lomax family.

In particular, the marginal density of $\mathbf x_S$, for $S\subset \{1,\ldots,d\}$, is
$\LM(|S|,k,a)$ which is based on the integral
$$\int_0^\infty (a+y)^{-k}\,dy=a^{-k+1}/(k-1)$$
for $k>1$. 
The conditional distribution of $\mathbf x_S$ given $\mathbf x_{T}$, for $S\cup T=\{1,\ldots,d\}$, is $\LM(|S|,k+d-|S|,a+\sum_{i\in T}x_i)$.

This is an important coherence condition: all the marginal and conditional prior distributions belong to the same family.  
For more on the multivariate Lomax distribution, see for example \cite{Hutchinson1979}, \cite{Lindley1986} and \cite{Nayak1987}.

\section{Illustrations}
In this Section we look at some implementations of the multivariate Lomax as an objective prior. In particular, we perform some single sample analysis as well as the analysis of the frequentist performance of the prior on repeated samples. For the latter, we evaluate the behaviour of the prior density with respect to the mean squared error (MSE), possibly in its more informative version as the relative square root MSE, and the coverage of the 95\% posterior credible intervals.

We first implement the proposed prior to estimate the parameters of a Weibull distribution, as an illustration with a parameter of dimension two.
We then present a key example to illustrate the actual implementation of the objective multivariate prior on a parameter space of dimension larger than two, by focusing on the three parameters of the Dagum distribution \citep{Dagum1975}.
Finally, we use the multivariate Lomax prior to estimate the parameters of a linear regression model, with normal error terms and two covariates. The prior is jointly assigned on the intercept, the two coefficients and the regression variance and, as far as we are aware, this is the sole objective prior distribution allowing this in a straightforward manner.

\subsection{Weibull distribution}
\label{sc_weibull}
Bayesian estimation of the parameters of a Weibull distribution using objective priors has been the object of some attention in the literature, with particular focus on Jeffreys prior and the reference prior. A discussion, with comparisons on the possible objective priors, can be found in \cite{Sun1997}. The author shows the superiority of the reference prior, although the choice of which (if any) parameter is nuisance is fundamental and gives rise to different priors.

We consider the Weibull distribution with the following parametrisation
$$f(x|\theta,\beta) = \beta x^{\beta-1}\,\theta^{-\beta}\exp\left\{-(x/\theta)^\beta\right\},\qquad x>0,$$
where $\theta$ is the scale parameter and $\beta$ the shape parameter. The objective prior against which we compare our proposal is the reference prior under the scenario where both parameters are of interest, that is
$\pi(\theta,\beta) \propto (\theta\beta)^{-1}$.
Following the guidelines of the simulation study in \cite{Sun1997}, we have drawn 250 independent samples of sizes $n=30$ and $n=100$, respectively, from a Weibull distribution with $\theta=1$ and $\beta=0.5, 1, 10$. This is due to the fact that changes in the scale parameter do not affect the behaviour of the prior.

\begin{table}[!ht]
    \centering
    \begin{tabular}{lccc|lccc}
    \hline
        MSE:  $\beta$ & ~ & ~ & ~ & MSE:  $\theta$ & ~ & ~ & ~ \\
        ~ & $\beta=0.5$ & $\beta=1$ & $\beta=10$ & ~ & $\beta=0.5$ & $\beta=1$ & $\beta=10$ \\
        Reference & 3.92 & 3.90 & 3.91 & ~ & 3.13 & 3.13 & 3.12 \\
        Lomax & 3.33 & 3.27 & 3.21 & ~ & 2.59 & 2.61 & 2.69 \\
        ~ & ~ & ~ & ~ & ~ & ~ & ~ & ~ \\ \hline
        COV: $\beta$ & ~ & ~ & ~ & COV: $\theta$ & ~ & ~ & ~ \\
        ~ & $\beta=0.5$ & $\beta=1$ & $\beta=10$ & ~ & $\beta=0.5$ & $\beta=1$ & $\beta=10$ \\
        Reference & 0.90 & 0.91 & 0.91 & ~ & 0.91 & 0.92 & 0.91 \\
        Lomax & 0.91 & 0.91 & 0.90 & ~ & 0.95 & 0.96 & 0.96 \\ \hline
    \end{tabular}
    \caption{Mean squared error and coverage of the 95\% posterior credible interval for the parameters of a Weibull, with $\theta=1$ and sample size $n=30$, based on 250 independent samples for each parameter combination.}
    \label{tab:weibulln30}
\end{table}

The results of the simulation study are summarised in Table \ref{tab:weibulln30}, for $n=30$, and in Table \ref{tab:weibulln100}, for $n=100$. Concerning the relative square root mean squared error (MSE), given by $\sqrt{\mbox{MSE}(\theta-\widehat{\theta})}/\theta$, where $\widehat{\theta}$ is the posterior mean, we note that in every scenario the Lomax prior outperforms the reference prior. Furthermore, as one would expect, for both priors the MSEs are smaller when the sample size is larger. In terms of coverage of the 95\% posterior credible intervals, we do not observe any particular deviation from what one would sensibly expect, as the values are within the normal range of tolerance.

\begin{table}[!ht]
    \centering
    \begin{tabular}{lccc|lccc}
    \hline
        MSE: $\beta$ & ~ & ~ & ~ & MSE: $\theta$ & ~ & ~ & ~ \\
        ~ & $\beta=0.5$ & $\beta=1$ & $\beta=10$ & ~ & $\beta=0.5$ & $\beta=1$ & $\beta=10$ \\
        Reference & 1.85 & 1.86 & 1.94 & ~ & 1.36 & 1.37 & 1.37 \\
        Lomax & 1.77 & 1.75 & 1.76 & ~ & 1.29 & 1.29 & 1.30 \\
        ~ & ~ & ~ & ~ & ~ & ~ & ~ & ~ \\ \hline
        COV: $\beta$ & ~ & ~ & ~ & COV: $\theta$ & ~ & ~ & ~ \\
        ~ & $\beta=0.5$ & $\beta=1$ & $\beta=10$ & ~ & $\beta=0.5$ & $\beta=1$ & $\beta=10$ \\
        Reference & 0.94 & 0.95 & 0.94 & ~ & 0.94 & 0.93 & 0.94 \\
        Lomax & 0.93 & 0.93 & 0.92 & ~ & 0.95 & 0.94 & 0.94 \\ \hline
    \end{tabular}
    \caption{Mean squared error and coverage of the 95\% posterior credible interval for the parameters of a Weibull, with $\theta=1$ and sample size $n=100$, based on 250 independent samples for each parameter combination.}
    \label{tab:weibulln100}
\end{table}

To conclude this study, analyze some real data, taken from \cite{Ellah2012}, and consisting of 19 times to breakdown (in minutes) of an insulating fluid between electrodes at a voltage of 34 KV (Table \ref{tab:weibullbrealdata}).

\begin{table}[!ht]
    \centering
    \begin{tabular}{cccccccccc}
    \hline
         0.96 & 4.15 & 0.19 & 0.78 & 8.01 & 31.75 & 7.35 & 6.50 & 8.27 & 33.91 \\ 
         32.52 & 3.16 & 4.85 & 2.78 & 4.67 & 1.31 & 12.06 & 36.71 & 72.89 & \\
         \hline
    \end{tabular}
    \caption{Observations of 19 times to breakdown (in minutes) of an insulating fluid between electrodes at a voltage of 34 KV.}
    \label{tab:weibullbrealdata}
\end{table}

\begin{figure*}[!ht]
        \centering
        \begin{subfigure}[b]{0.475\textwidth}
            \centering
            \includegraphics[width=\textwidth]{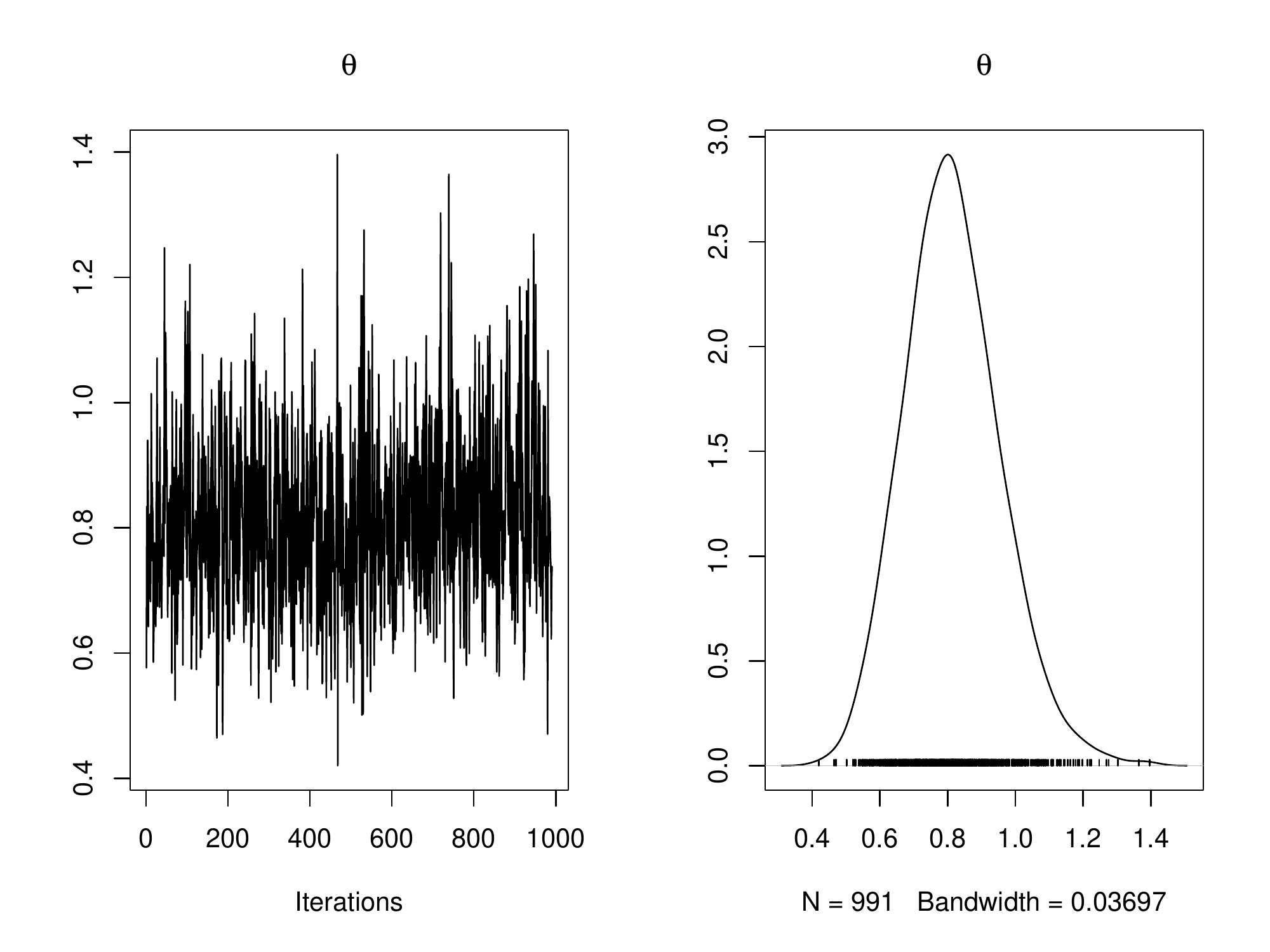}
            \caption[]%
            {{\small Reference prior: $\theta$}}    
            \label{fig:mean and std of net14}
        \end{subfigure}
        \hfill
        \begin{subfigure}[b]{0.475\textwidth}  
            \centering 
            \includegraphics[width=\textwidth]{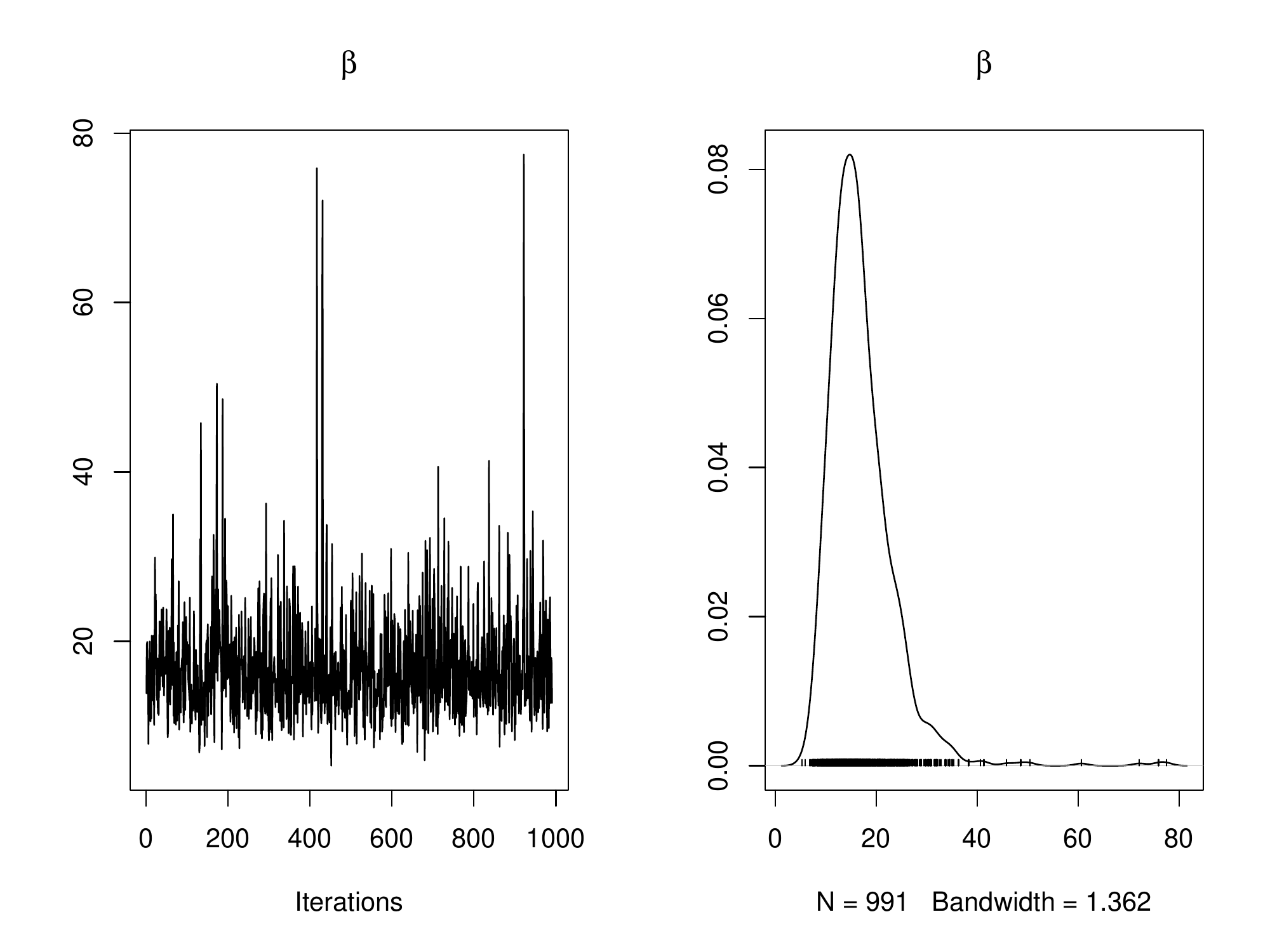}
            \caption[]%
            {{\small Reference prior: $\beta$}}    
            \label{fig:mean and std of net24}
        \end{subfigure}
        \vskip\baselineskip
        \begin{subfigure}[b]{0.475\textwidth}   
            \centering 
            \includegraphics[width=\textwidth]{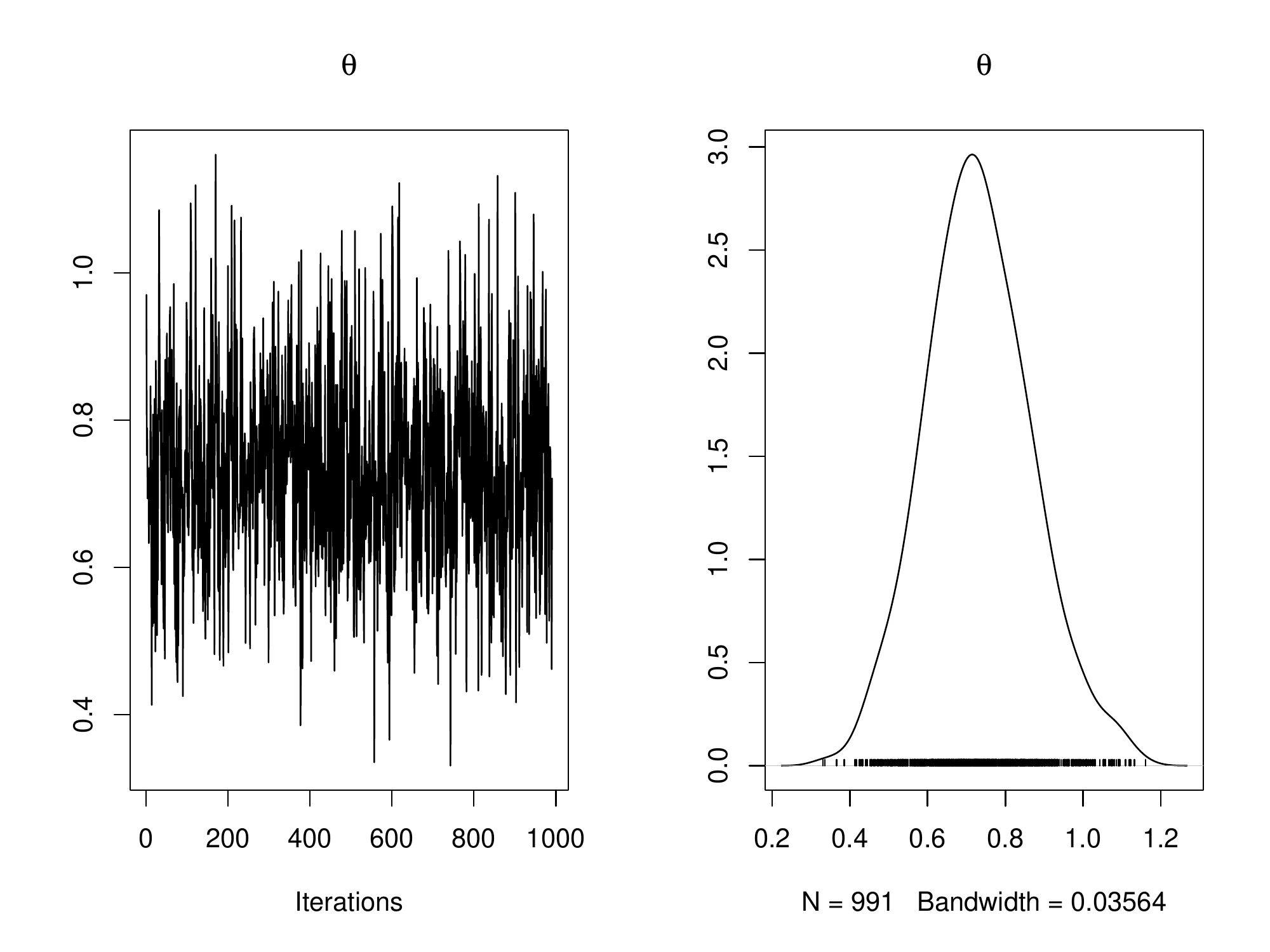}
            \caption[]%
            {{\small Lomax prior: $\theta$}}    
            \label{fig:mean and std of net34}
        \end{subfigure}
        \hfill
        \begin{subfigure}[b]{0.475\textwidth}   
            \centering 
            \includegraphics[width=\textwidth]{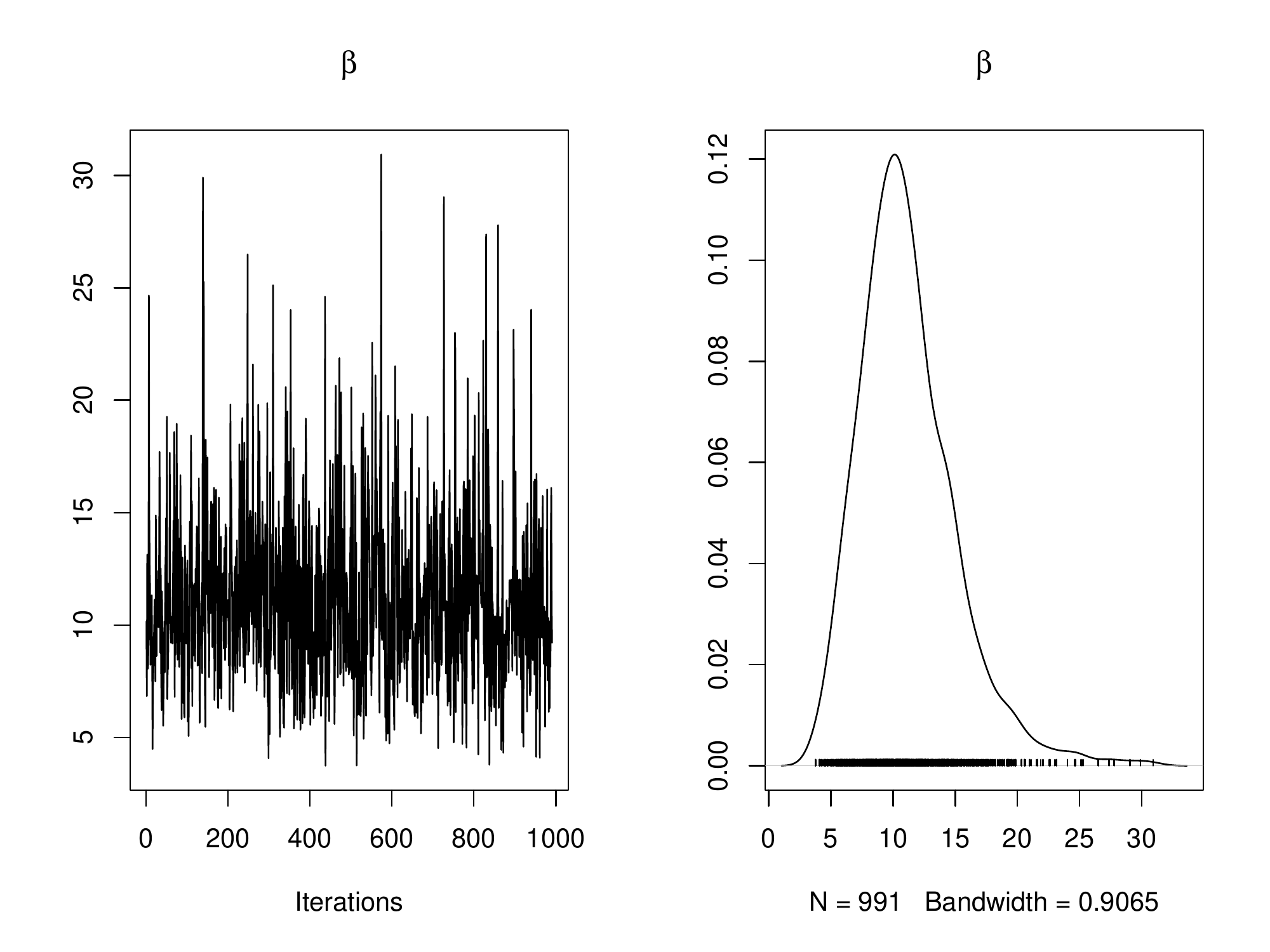}
            \caption[]%
            {{\small Lomax prior: $\beta$}}    
            \label{fig:mean and std of net44}
        \end{subfigure}
        \caption[]
        {\small Posterior chains and estimated densities for the scale (left) and shape (right) parameters of the Weibull using the reference prior (top) and the Lomax prior (bottom).} 
        \label{fig:weibullrealdata}
    \end{figure*}

The maximum likelihood estimates of the parameters are $\widehat{\theta}=0.77$ and $\widehat{\beta}=12.22$. 
Looking at the posterior statistics of the posteriors under the reference prior and the Lomax prior (Table \ref{tab:weibullrealdata}), we see that the intervals obtained with both priors include MLE estimates of the corresponding parameters. For $\theta$, the performance of the two priors is quite similar. For the shape parameter $\beta$, however, we see that the Lomax prior outperforms the reference prior, which is obvious by comparing both the posterior variances and 95\% credible intervals.
\begin{table}[!ht]
\begin{tabular}{ccccccc}
\hline
\multicolumn{1}{l}{} & \multicolumn{3}{c}{Reference}   & \multicolumn{3}{c}{Lomax}       \\
                     & Mean  & Variance & 95\% C.I.           & Mean  & Variance & 95\% C.I.           \\ \hline
$\theta$                & 0.8   & 0.02     & (0.55,1.10)  & 0.73  & 0.02     & (0.48,1.02)  \\
$\beta$                 & 16.84 & 44.93    & (8.51,31.89) & 11.11 & 15.08    & (5.07,20.36)\\
\hline
\end{tabular}
\caption{Posterior summaries for the Weibull real data example. The MLE are $\widehat{\theta}=0.77$ and $\widehat{\beta}=12.22$.}
\label{tab:weibullrealdata}
\end{table}
To complete this illustration, we show the posterior chains and estimated densities in Fig.~\ref{fig:weibullrealdata}.

\subsection{Dagum distribution}
\label{sc_dagum}
The Dagum distribution was introduced for the first time in \cite{Dagum1975}, and is widely used in economics, for example, to model income and wealth data. We say that a random variable $X$ follows a Dagum distribution if it has the following probability density function:
$$f(x|a,b,p) = \frac{a\,p}{x}\left(\dfrac{\left\{x/b\right\}^{ap}}{\left\{(x/b)^a+1\right\}^{p+1}}\right),\qquad x>0,$$
where $p>0$ and $a>0$ are shape parameters, and $b>0$ is a scale parameter.

The Bayesian literature on adopting objective priors to estimate the parameters of a Dagum distribution is scarce, with the exception of \cite{NAA2017}, where the authors propose an extension of Jeffreys prior; however, the proposed prior is not, strictly speaking, an objective prior as it depends on a hyper-parameter that has to be set and which impacts the performance of the prior itself. As such, in order to have a suitable comparison of the multivariate Lomax with an alternative solution, we will consider a joint prior on the three parameters, to be the product of three independent vague gamma densities. That is, $\pi(p,a,b)=\mbox{Ga}(a_1,b_1)\cdot \mbox{Ga}(a_2,b_2)\cdot \mbox{Ga}(a_3,b_3)$, with $a_1$, $a_2$, $a_3$, $b_1$, $b_2$ and $b_3$ relatively small (e.g. equal to 0.1 or 0.01).

To compare the multivariate Lomax prior to the above vague prior, we analyse two scenarios. First, we consider a Dagum distribution with parameters $(p=1, a=2.1, b=1)$, and then a Dagum with parameters $(p=2, a=2.1, b=2)$. We have chosen $a>2$ so that both the mean and the variance of the distribution exist. We replicate 250 independent samples for each scenario with sizes $n=30$ and $n=100$. The posterior distributions for the parameters are not available in closed form, so we implemented a Metropolis within Gibbs to sample from them.

In Tables \ref{tab:dagumfirst} and \ref{tab:dagumsecond}, we repor the relative (square root) mean squared error and coverage of the 95\% posterior credible intervals for the above scenarios, under the two priors.
\begin{table}[h!]
\centering
\begin{tabular}{l|cccc}
\hline
\textbf{MSE}       & Vague        & Lomax       & Vague        & Lomax        \\
Parameter & \multicolumn{2}{c}{n = 30} & \multicolumn{2}{c}{n = 100} \\
\hline
$p$         & 13.28        & 6.59        & 7.87         & 4.89         \\
$a$         & 2.80         & 2.05        & 1.45         & 1.24         \\
$b$         & 4.26         & 3.18        & 2.92         & 2.44         \\
\hline
\textbf{COV}       & Vague        & Lomax       & Vague        & Lomax        \\
Parameter & \multicolumn{2}{c}{n = 30} & \multicolumn{2}{c}{n = 100} \\
\hline
$p$         & 0.97         & 1.00        & 0.95         & 0.97         \\
$a$         & 0.96         & 0.99        & 0.96         & 0.96         \\
$b$         & 0.98         & 1.00        & 0.94         & 0.96        \\
\hline
\end{tabular}
\caption{Relative MSE (top) and Coverage of the 95\% posterior credble interval (bottom) for the parameters of a Dagum distribution, based on 250 repeated samples of sizes $n=30$ and $n=100$, with $p=1$, $a=2.1$ and $b=1$. The tables compares values obtained using the multivariate Lomax and a product of three independent vague Gamma densities.}
\label{tab:dagumfirst}
\end{table}
We notice that, in both cases, the multivariate Lomax outperforms the joint vague Gamma density in terms of the MSE. This appears to be more prominent for the estimates of the parameter $p$, suggesting a better accuracy of the Lomax prior. Regarding coverage, the priors tend to behave similarly, with relatively larger intervals for the smaller sample sizes, as one would expect.
\begin{table}[h!]
\centering
\begin{tabular}{l|cccc}
\hline
\textbf{MSE}       & Vague        & Lomax       & Vague        & Lomax        \\
Parameter & \multicolumn{2}{c}{n = 30} & \multicolumn{2}{c}{n = 100} \\
\hline
$p$         & 8.05         & 3.55        & 9.21         & 4.52         \\
$a$         & 2.38         & 2.03        & 1.24         & 1.11         \\
$b$         & 3.96         & 2.80        & 3.09         & 2.38         \\
\hline
\textbf{COV}       & Vague        & Lomax       & Vague        & Lomax        \\
Parameter & \multicolumn{2}{c}{n = 30} & \multicolumn{2}{c}{n = 100} \\
\hline
$p$         & 0.99         & 0.99        & 0.97         & 0.98         \\
$a$         & 0.96         & 0.98        & 0.97         & 0.98         \\
$b$         & 1.00         & 1.00        & 0.95         & 0.98        \\
\hline
\end{tabular}
\caption{Relative MSE (top) and Coverage of the 95\% posterior credble interval (bottom) for the parameters of a Dagum distribution, with $p=2$, $a=2.1$ and $b=2$, based on 250 repeated samples of sizes $n=30$ and $n=100$. The tables compares values obtained using the multivariate Lomax and a product of three independent vague Gamma densities.}
\label{tab:dagumsecond}
\end{table}

\begin{figure}[h!]
    \centering
    \includegraphics[width=\textwidth]{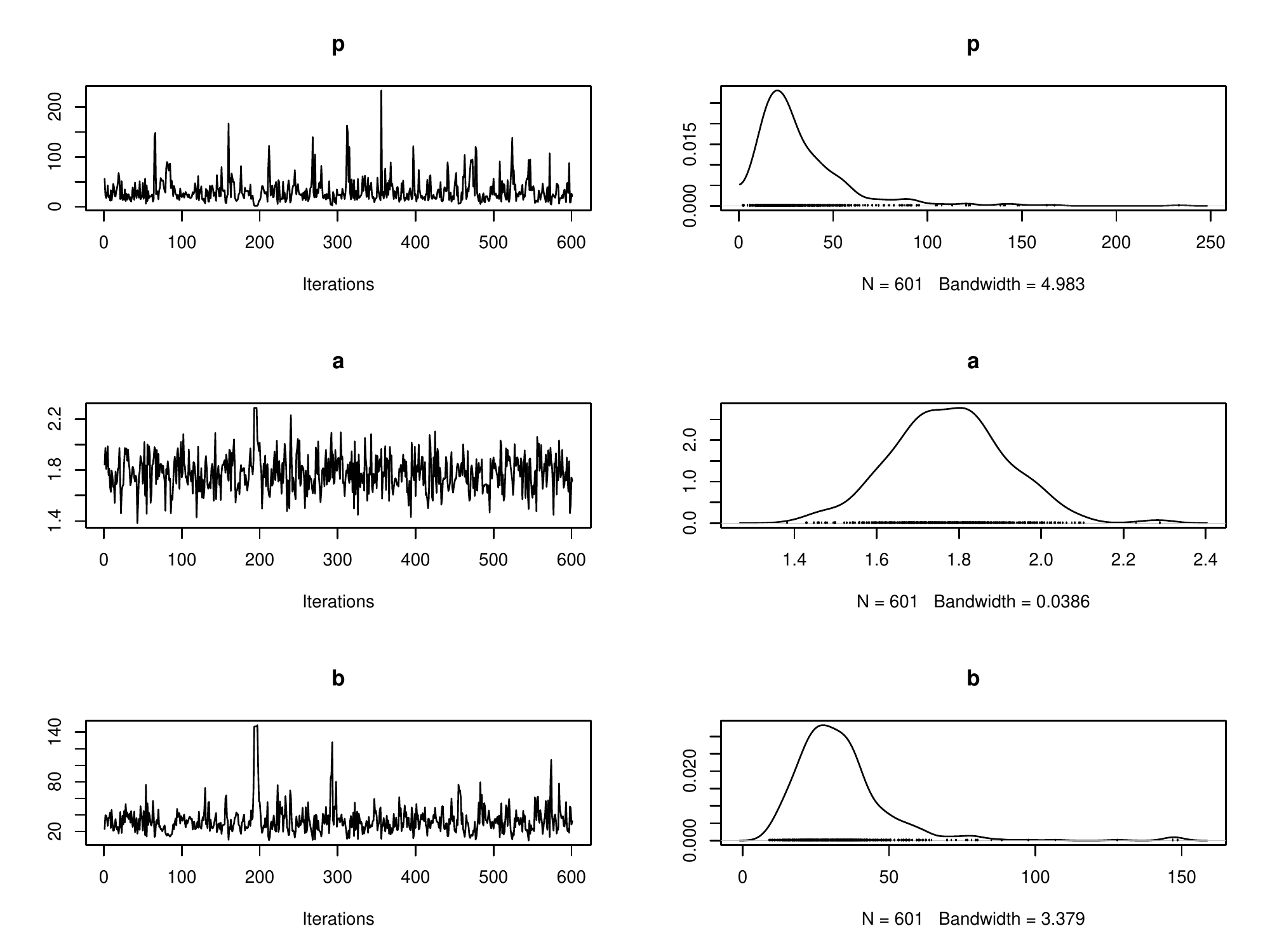}
    \caption{Posterior chains and estimated densities for the parameters of the Dagum using the multivariate lomax prior.}
    \label{fig:dagumlomax}
\end{figure}

\begin{figure}[h!]
    \centering
    \includegraphics[width=\textwidth]{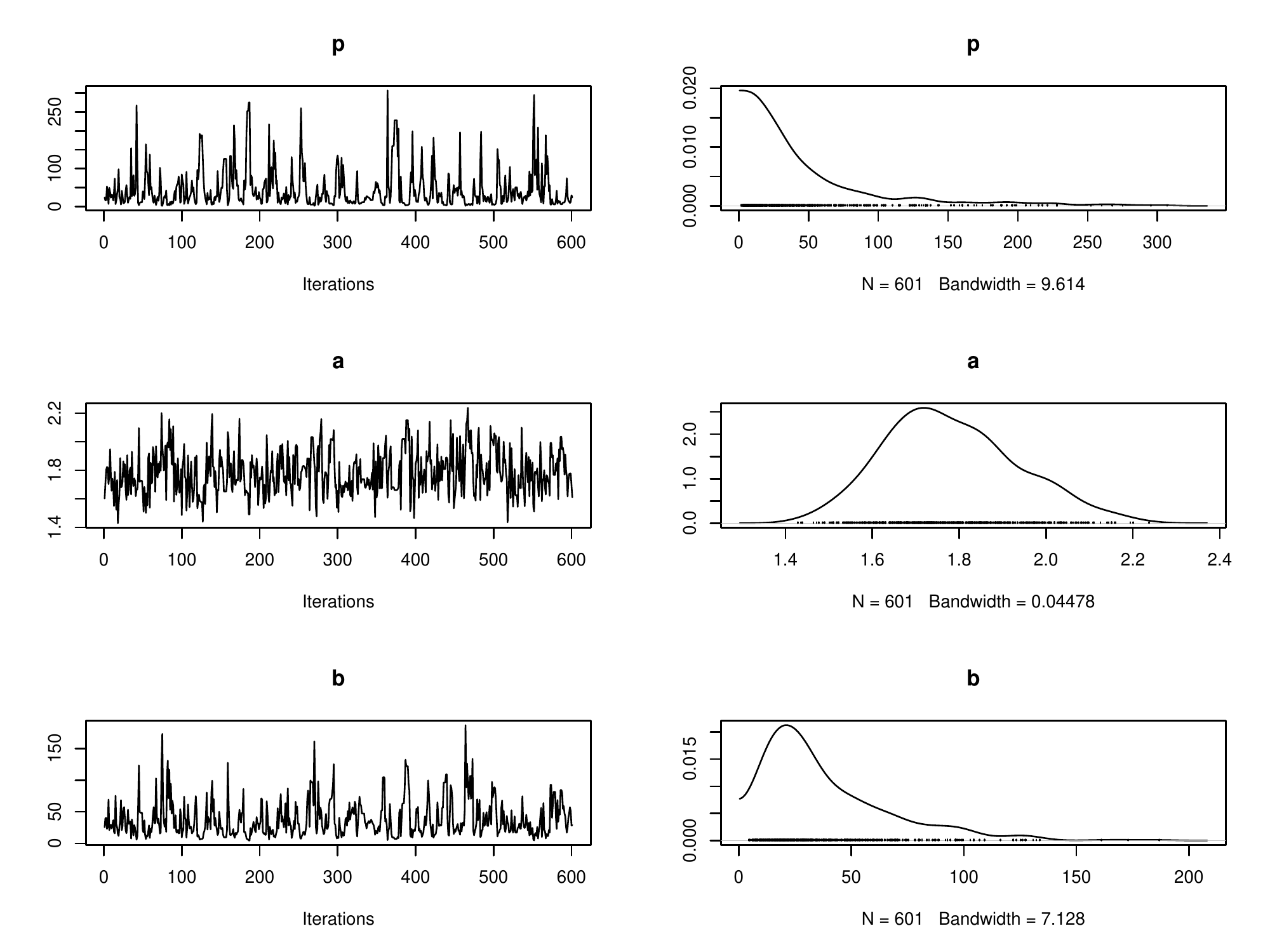}
    \caption{Posterior chains and estimated densities for the parameters of the Dagum using independent vague Gamma priors.}
    \label{fig:dagumgamma}
\end{figure}

To conclude the illustration of the Dagum density, we analyse the UK Quarterly Gas Consumption in the period 1960--1986 \citep{Durbin2001}, a publicly available data set. We compare the thre-dimensional multivariate Lomax with the product of three vague Gamma densities.
In Figure \ref{fig:dagumlomax}, we have the posterior traces and estimated densities that have been obtained when the multivariate Lomax has been used as joint prior on the three parameters of the Dagum distribution, assumed as statistical model for the data. In Figure \ref{fig:dagumgamma}, we have the same plots when a product of three independent vague Gamma densities has been used as joint prior. It is clear that, while the three marginal posterior densities appear to be centred around the same values under both priors, the variability, and so the uncertainty, expressed by the posteriors under the Lomax prior seems to be smaller than the competing vague Gamma density. This is confirmed by the posterior statistics reported in Table \ref{tab:dagumrealdata}.

\begin{table}[h!]
\begin{tabular}{c|cc|cc}
Parameter & \multicolumn{2}{c}{Lomax} & \multicolumn{2}{c}{Gamma} \\
\hline
p         & 25.5    & (6.61, 104.19)  & 27.2   & (3.40, 209.14)   \\
a         & 1.77    & (1.50, 2.04)    & 1.76   & (1.52, 2.10)     \\
b         & 30.66   & (12.86, 76.68)  & 29.1   & (7.31, 109.23)  
\end{tabular}
\caption{Posterior median and posterior 95\% credible intervals for the parameters of the Dagum in a real data example. To the left, the values obtained using a multivariate Lomax prior, while to the right we have the values obtained using a prior formed by the product of three independent vague Gamma densities.}
\label{tab:dagumrealdata}
\end{table}

\subsection{Linear regression}\label{sc_linearreg}
An interesting application of the proposed prior  is to estimate the parameters of a linear regression model. Typically, objective prior distributions for this type of model separate the coefficients from the intercept and the regression variance. In other words, for the model 
$$y_i=\beta_0+\sum_{i=1}^px_{ij}\beta_j+\varepsilon_i,\quad \mbox{for}\quad i=1,\ldots,n,$$ 
with $\varepsilon_i\sim N(0,\sigma^2)$, the objective prior would be of the type 
$$\pi(\beta_0,\beta_1,\ldots,\beta_p,\sigma)=\pi(\beta_0,\sigma)\pi(\beta_1,\ldots,\beta_p).$$ 
Usually, in an objective Bayesian set up, one would set $\pi(\beta_0,\sigma)\propto\sigma^{-1}$, while the prior on the coefficients can take different forms, such as in \cite{ZS1980}, \cite{Fernandez2001}, \cite{Liang2008} and \cite{Bayarri2012}.
Here, we discuss the implementation of the proposed prior via a single sample example and a frequentist analysis performed on a repeated sampling scheme.

\subsubsection{Linear regression: Single sample}\label{sc_regsinglesample}
Let us consider a sample of size $n=100$ drawn from a linear regression model with two covariates, $x_1$ and $x_2$ (both defined on $\mathbb{R}$), intercept $\beta_0=20$, coefficients $\pmb\beta=(10,-1)$, and variance $\sigma^2=2$. The (4-dimensional) Lomax prior has the following form,
\begin{equation}\label{eq:regdim4}
\pi(\beta_0,\beta_1,\beta_2,\sigma^2) = 3\,\left(1+\sum_{j=0}^2|\beta_j|+\sigma^2\right)^{-5},
\end{equation}
with $\beta_j\in\mathbb{R}$, for $j=0,1,2$, and $\sigma^2>0$. To study the posterior distribution, we implemented a Markov Chain Monte Carlo (MCMC) algorithm with 100,000 iterations, taking a burn--in of 10,000 samples and a thinning gap of 50 samples.

\begin{figure}[!ht]
    \centering
    \includegraphics[width=12cm, height=12cm]{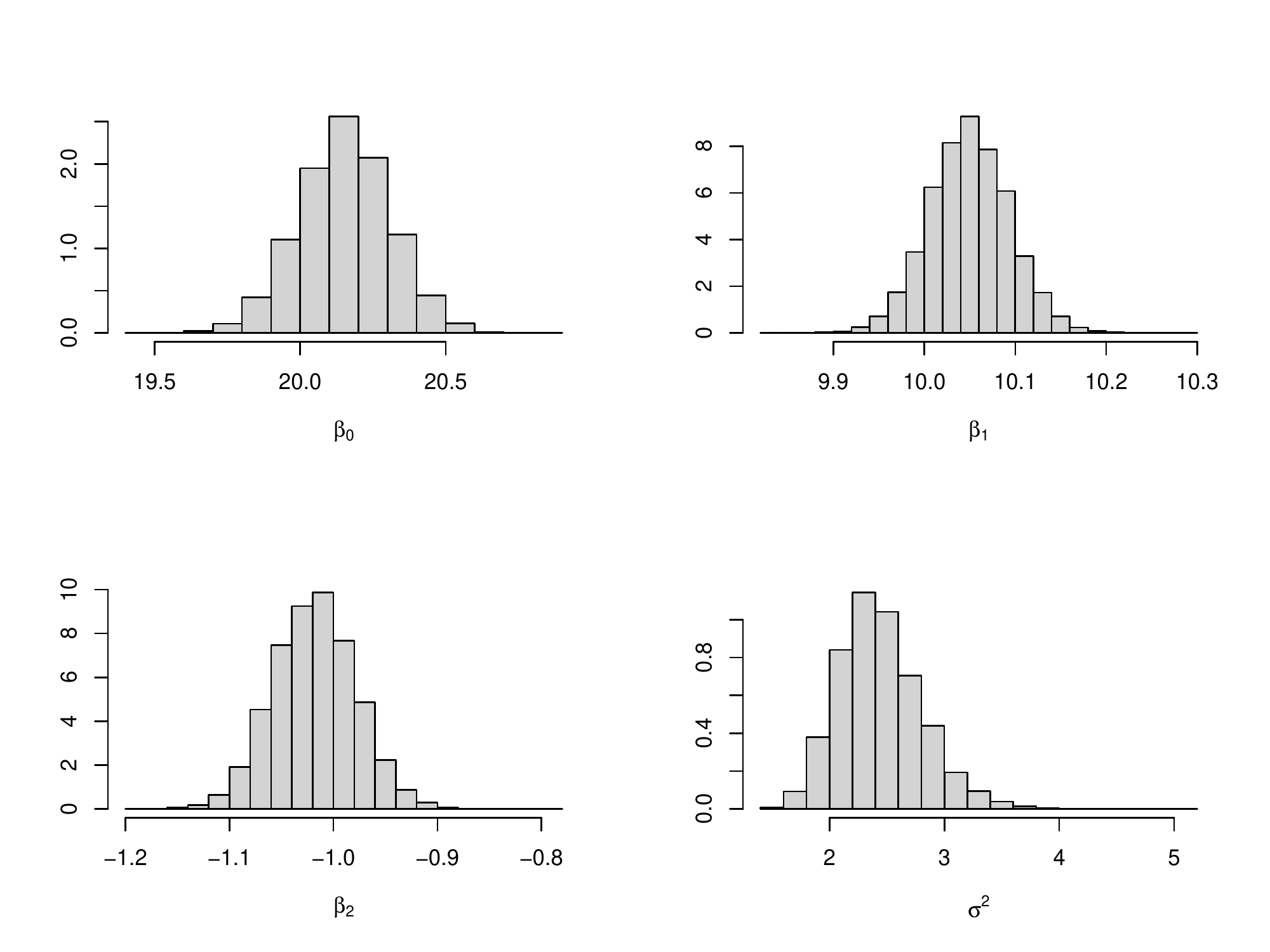}
    \caption{Histograms of the posterior samples for the parameters pf the linear regression model with intercept $\beta_0=20$, coefficients $\beta_1=10$ and $\beta_2=-1$, and variance $\sigma^2$=2}.
    \label{fig:regression_singlesample}
\end{figure}
In Fig.~\ref{fig:regression_singlesample} we present the histograms of the four marginal posterior samples, while in Table \ref{tab:reg_singlesample} we report the corresponding posterior summary statistics.
\begin{table}[htbp]
    \centering
    \begin{tabular}{c|cc}
        Parameter & Median & 95\% C.I. \\
        \hline
        $\beta_0$ & 20.15 & (19.84, 20.47)\\
        $\beta_1$ & 10.05 & (9.96, 10.14)\\
        $\beta_2$ & -1.02 & (-1.10, -0.94)\\
        $\sigma^2$ & 2.41 & (1.83, 3.26)
    \end{tabular}
    \caption{Posterior summary statistics for the single-sample regression example.}
    \label{tab:reg_singlesample}
\end{table}
We see that the true parameter values are well contained in the posterior 95\% credible intervals, showing a correct inferential performance of the proposed prior.

\subsubsection{Linear regression: Frequentist analysis}\label{sc_regfrequentist}
To show the performance of the proposed prior, and to compare it to other methods, we have drawn 250 samples of size $n=100$ from a linear regression model with intercept $\beta_0=20$, coefficients $\beta_1=5$ and $\beta_2=-3$, and variance $\sigma^2=2$.
The proposed prior is given in \eqref{eq:regdim4}, and it has been compared with an independent vague prior, that is
\begin{eqnarray*}
\pi(\beta_0,\beta_1,\beta_2,\sigma^2) &=& \pi(\beta_0,\beta_1,\beta_2)\pi(\sigma^2) \\
&\propto& \sigma^{-1}\,\mbox{MVN}(\mathbf{0},c\mbox{I}_3),
\end{eqnarray*}
where $\mbox{I}_3$ is a diagonal matrix of dimension 3, and $c>0$ has been chosen to be sufficiently large to represent vague prior information. Additionally, to compare the proposed prior to a prior that is considered {\em objective}, we have used the following version of Zellner's $g$ prior \citep{ZS1986}
\begin{eqnarray*}
\pi(\beta_0,\beta_1,\beta_2,\sigma^2) &=& \pi(\sigma)\pi(\beta_0,\beta_1,\beta_2|\sigma)\\
&\propto& \sigma^{-1}\,\mbox{MVN}(\mathbf{0},\sigma^2g\mathbf{X}),
\end{eqnarray*}
where $g>0$ has been calibrated to be 500, and $\mathbf{X}$ is the design matrix.
In Table \ref{tab:MSE100} we have reported the (square root) relative squared error from the maximum a posteriori (MAP) under each of the three prior distributions. It is easy to see that the values are virtually the same.
\begin{table}[h]
    \centering
    \begin{tabular}{c|ccc}
        Parameter & Lomax & Vague & Zellner's $g$ \\
        \hline
        $\beta_0$ & 0.998 & 0.998 & 0.998 \\
        $\beta_1$ & 1.000 & 1.000 & 1.000 \\
        $\beta_2$ & 1.001 & 1.002 & 1.001 \\
        $\sigma^2$ & 1.098 & 1.095 & 1.094 \\
    \end{tabular}
    \caption{Squared error form the MAP for the linear regression model, comparing the proposed prior to an independent vague prior and Zellner's $g$ prior.}
    \label{tab:MSE100}
\end{table}
In Table \ref{tab:COV100} we report the coverage of the 95\% posterior credible intervals under each prior. Again, the performances of the prior distributions are almost identical.
\begin{table}[htbp]
    \centering
    \begin{tabular}{c|ccc}
        Parameter & Lomax & Vague & Zellner's $g$ \\
        \hline
        $\beta_0$ & 0.95 & 0.94 & 0.95 \\
        $\beta_1$ & 0.97 & 0.96 & 0.97 \\
        $\beta_2$ & 0.97 & 0.98 & 0.98 \\
        $\sigma^2$ & 0.93 & 0.92 & 0.92 \\
    \end{tabular}
    \caption{Coverage of the 95\% posterior credible intervals for the linear regression model, comparing the proposed prior to an independent vague prior the Zellner's $g$ prior.}
    \label{tab:COV100}
\end{table}

\section{Discussion}\label{sc_discussion}

In this paper, we have presented a general approach to the specification of a multidimensional objective prior distribution. The prior is proper, as it is a multivariate Lomax distribution, and has some appealing coherency properties consisting on the fact that all joint, marginal and conditional distributions are also Lomax.

Through the analysis of both simulated and real data, we have showed how the proposed prior behaves in different scenarios, including when the data is assumed to be distributed as a Weibull or a Dagum, as well as when a linear regression model is considered. In all cases, the multivariate prior performs as well as the competing solutions and, for the Weibull and the Dagum cases, it actually outperforms the alternative priors.

An interesting final remark related to linear regression models, is that the multivariate Lomax on the coefficients produces a result that can be compared with the LASSO regression. In fact, it can be noted that the multivariate Lomax is a mixture of independent Laplace densities with respect to the exponential distribution. That is,
\begin{equation}\label{eq:laplace}
p(x_1,...,x_d) = \int \prod_{j=1}^d \mbox{Lap}(x_j | s) e^{-s} \, ds,
\end{equation}
where $\mbox{Lap}(\cdot|s)$ is a Laplace density with location zero and scale parameter $s$. If we apply the above \eqref{eq:laplace} to a lineal regression model, we would then need to minimise the following function
$$\sum_{i=1}^n (y_i-x_i \beta)^2+(d+1) \log(1+|\beta|),$$
which can be compared to the LASSO minimising function, for which the last term would be $|\beta|$, instead of $\log(1 + |\beta|))$. The idea is that having the penalty term in the log scale, would result in more robust estimates. More on this can be found in Appendix \ref{sec:lasso}.

\bibliography{MLomaxPrior.bib}

\appendix
\appendixpage

\section{Mathematical details}\label{sec:details}

Define $\mathbf x=(x_1,\ldots,x_d)$ and $\mathbf q = (q,q_1,\ldots,q_d)$ where, for $j=1,\ldots, d$
$$q_j=\frac{\partial q}{\partial x_j},
$$
and likewise for $\mathbf p$.
We consider the divergence between probability measures $P$ and $Q$ to be the Bregman divergence (see \cite{BREGMAN1967} and \cite{Ovcharov2018}):
\begin{equation}\label{eq:BDgen}
    D(P,Q) = \int_\Omega B_\phi(\bm{p},\bm{q})\,d\mathbf x,
\end{equation} 
where $p$ and $q$ are the densities associated with $P$ and $Q$, respectively, and
\begin{equation}\label{eq:Bgen-d}
    B_{\phi}(\bm{p},\bm{q}) = \phi(\bm{p})-\phi(\bm{q})-\frac{\partial\phi}{\partial q}(p-q)-\sum_{i=1}^d\frac{\partial\phi}{\partial q_i}(p_i-q_i),
\end{equation}
for some continuously differentiable convex function $\phi$ on $\mathbb R^{d+1}$.

\begin{lemma}
The corresponding Bregman score function is given by
$$S(\bm{q}) = -\frac{\partial\phi}{\partial q}+\sum_{i=1}^d\frac{\partial}{\partial x_i}\frac{\partial\phi}{\partial q_i},$$
under the assumption that the extremum values for
$$p_i(\mathbf x)\,\frac{d}{dx_i}\,\frac{\partial \phi}{\partial q_i}$$
disappear for each $i.$ and $\phi$ satisfies
\begin{equation}\label{phicond}
\phi(\mathbf q)=q\frac{\partial\phi}{\partial q}+\sum_{i=1}^d q_i\frac{\partial\phi}{\partial q_i}.
\end{equation}
\end{lemma}
\begin{proof}
The condition in the statement of the Lemma allows for an integration by parts of $p_i\,\partial\phi/\partial q_i$ for each $i$.
Hence, we can rewrite \eqref{eq:BDgen} as
$$\int\left\{\phi(\bm{p})+p\left[-\frac{\partial\phi}{\partial q}+\sum_{i=1}^d\frac{\partial}{\partial x_i}\frac{\partial\phi}{\partial q_i}\right] - \left[\phi(\bm{q})-q\frac{\partial\phi}{\partial q}-\sum_{i=1}^d q_i\frac{\partial\phi}{\partial q_i}\right]\right\}\, d\mathbf x.$$
which identifies the score function as given, with $\phi$ satisfying the constraint provided in the statement of the Lemma
\end{proof}

To meet the constraint for $\phi$ we define
$\alpha:\mathbb R_+^d\rightarrow \mathbb R$ to be
$$\alpha(\mathbf u)=\sum_{i=1}^d u_i^{-(k+d)}$$ 
and $u_i=q_i/q$ for any density $q$ on $(0,\infty)$. 
Note that the first and second partial derivatives are
$$\alpha_i(\mathbf u)=-(k+d)u_i^{-(k+d+1)}\quad\mbox{and }
\alpha_{ii}(\mathbf u)=(k+d)(k+d+1)u_i^{-(k+d+2)},
$$ 
so the Hessian matrix has determinant 
$$|H\alpha(\mathbf u)|=(k + d)^d (k + d+1)^d \prod_{i=1}^d u_i^{-(k + d+2)}$$ 
and therefore $\alpha(\mathbf u)$ is convex for $u\in (0,\infty)^d$  whenever $k\in\mathbb N$ is odd.
\\

Let 
\begin{equation}\label{eq:proposedphi-d}
\phi(\mathbf q) = q\,\alpha\left(\frac{q_1}{q},\ldots,\frac{q_d}{q}\right).
\end{equation}

\begin{lemma}
The function $\phi$ in equation \eqref{eq:proposedphi-d} satisfies (\ref{phicond}).
\end{lemma}
\begin{proof} We present the proof in the case of $d=2$, for higher dimensions the math follows likewise. In the following $q_1=q_x$ and $q_2=q_y$. Now 
\begin{align*}\label{eq:check}
    \frac{\partial\phi}{\partial q} &= q\left[\alpha_u\left(\frac{q_x}{q},\frac{q_y}{q}\right)\left(\frac{q_x}{q^2}\right) + \alpha_v\left(\frac{q_x}{q},\frac{q_y}{q}\right)\left(-\frac{q_y}{q^2}\right)\right] + \alpha\left(\frac{q_x}{q},\frac{q_y}{q}\right) \nonumber\\
    &= \alpha\left(\frac{q_x}{q},\frac{q_y}{q}\right)-\frac{1}{q}\left[q_x\alpha_u\left(\frac{q_x}{q},\frac{q_y}{q}\right)+q_y\alpha_v\left(\frac{q_x}{q},\frac{q_y}{q}\right)\right],
\end{align*}
and
$$\frac{\partial\phi}{\partial q_x} = q\,\alpha_u\left(\frac{q_x}{q},\frac{q_y}{q}\right)\frac{1}{q} = \alpha_u\left(\frac{q_x}{q},\frac{q_y}{q}\right),$$
and
$$\frac{\partial\phi}{\partial q_y} = \alpha_v\left(\frac{q_x}{q},\frac{q_y}{q}\right),$$
showing that condition \eqref{phicond} is satisfied. 
\end{proof}

\begin{corollary}
The score-based prior for a parameter $\mathbf x\in\mathbb R_+^d$, associated to the Bregman divergence \eqref{eq:Bgen-d} with convex function $\phi$ given by  \eqref{eq:proposedphi-d}, for  $\alpha(\mathbf u)=\sum_{i=1}^d u_i^{-(k+d)}$ is the multivariate Lomax distribution $\LM(d,a,k)$.
\end{corollary}

\section{More on the Alternative to LASSO}\label{sec:lasso}

Here we discuss the idea of regularization using the penalty term $\log(1+|\beta|)$ for variable selection, so the aim is to minimize
$$\sum_{i=1}^n (y_i-x_i'\beta)^2+2\lambda \log(1+|\beta|),$$
for some $\lambda>0$. It is easier to see the consequences of using $\log(1+|\beta|)$, rather than the more usual $|\beta|$ which characterizes the LASSO, if we move to one dimension, for which the problem reduces to minimizing
$$\beta^2-2\beta z+2\lambda \log(1+|\beta|).$$
Strictly, this is assuming that $\sum_{i=1}^n x_i^2=1$ and $z=\sum_{i=1}^n x_i\,y_i$.

The solution to this is easy to find:
$$\widehat{\beta}=\left\{
\begin{array}{ll}
0 & z<\lambda \\
(z-1)/2+ \sqrt{z-\lambda+(z-1)^2/4} & z>\lambda,
\end{array}\right.
$$
if $z>0$ and
$$\widehat{\beta}=\left\{
\begin{array}{ll}
0 & z>-\lambda \\
(1+z)/2- \sqrt{-\lambda-z+(1+z)^2/4} & z<-\lambda,
\end{array}\right.
$$
if $z<0$.

In many aspects this solution is similar to the LASSO problem, i.e. to minimize
$$\beta^2-2\beta z+2\lambda |\beta|,$$
for which the solution is $\widehat{\beta}=0$ for $|z|<\lambda$, and is $\widehat{\beta}=z-\lambda$ for $z>\lambda$ and $\widehat{\beta}=z+\lambda$ for $z<-\lambda$. Hence, as $z\to \infty$, for example, the LASSO estimator never returns to the standard unbiased estimator $z$. However, for the $\log(1+|\beta|)$ penalty it does. We see that as $z\to\infty$, so
$$(z-1)/2+ \sqrt{z-\lambda+(z-1)^2/4}- z\to 0.$$
It is important to note that both the LASSO and the log penalty achieve $\widehat{\beta}=0$ for $|z|<\lambda$. Hence, it is when $|z|>\lambda$ that we should focus on, and the simple fact here is that for $|z|>\lambda$ the log penalty estimator is close to $z$ which itself is closer to the true parameter. 

\begin{figure}[h!]
    \centering
    \includegraphics[width=12cm,height=6cm]{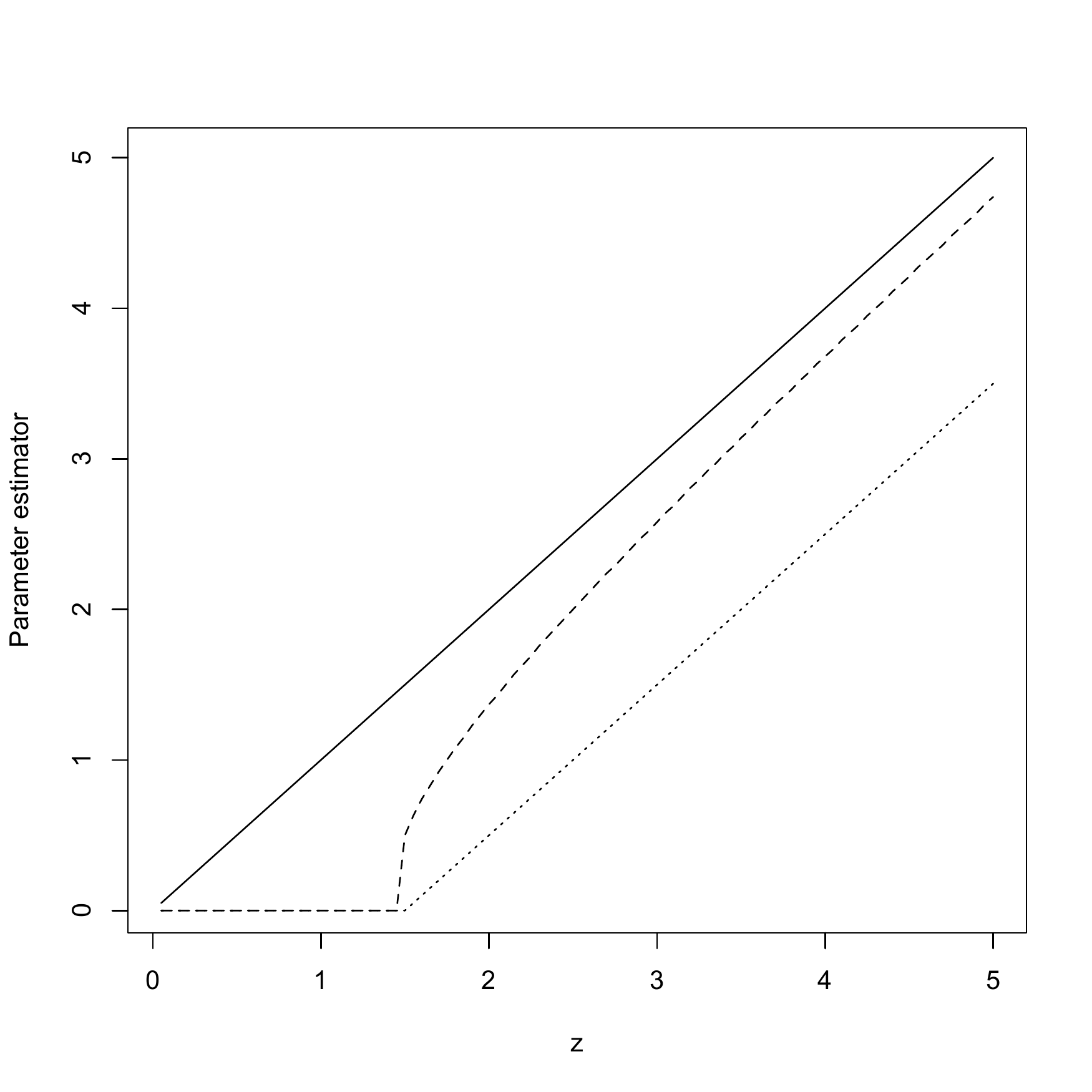}
    \caption{Ordinary least square estimator, bold line, along with LASSO estimator, dotted line, and log penalty estimator, dashed line. }
    \label{figL}
\end{figure}

To conduct such an experiment, we took $n=100$ and sampled 
$$y_i=\beta+\varepsilon_i,$$
where the $(\varepsilon_i)$ are independent standard Normal random variables. Computing $z=\bar{y}$, we obtain the estimators based on the LASSO, $\widehat{\beta}_L$, and the log penalty, $\widehat{\beta}_N$, and over a repeat of such an experiment 1000 times we get a mean square error for each estimator; i.e.
$M_L=E(\widehat{\beta}_L-\beta)^2$ and $M_N=E(\widehat{\beta}_N-\beta)^2$. Every scenario results in a smaller $M_N$ compared with $M_L$. For example, with $\lambda=1/2$ and $\beta=2$, we obtain $M_L=0.260$ and $M_N=0.043$. Keeping $\lambda=1/2$ and now setting $\beta=1/2$ we get $M_L=0.216$ and $M_N=0.197$. These improvements are based on the fact that for $z>\lambda$ we have $\widehat{\beta}_L<\widehat{\beta}_N<z$ and for $z<-\lambda$ it is that $z<\widehat{\beta}_N<\widehat{\beta}_L$. See Fig.~\ref{figL}.

With $\beta=0$ we get $M_L=M_N=0$ indicating $|z|<\lambda$ in every case. If we keep $\beta=0$ and now take $\lambda=0.1$ we get $M_L=0.0016$ and $M_N=0.0019$.
This latter result is easy to explain due to the closer the LASSO estimator is to 0 when $|z|>\lambda$. However, when $\beta\ne 0$ the log penalty estimator provides a smaller mean square error. 

\end{document}